\documentstyle[12pt]{article}
\begin{document}
\hoffset=-0.3in
\voffset=-0.8in
\vspace*{2cm}
\begin{center}
{\large\bf The explanation of unexpected temperature dependence of the muon
catalysis in solid deuterium}
\end{center}

\vspace*{1cm}
\begin{center}
S.S. Gershtein\footnote{E-mail: gershtein@mx.ihep.su},\\
{\it Institute for High Energy Physics, Protvino, Moscow  region, 142284
Russia}
\end{center}

\vspace*{1.cm}
\underline{Abstract}

\small
\noindent
It is shown that due to the smallness of the inelastic cross-section of the
$d\mu$-atoms scattering in the crystal lattice  at sufficiently low
temperatures the $dd\mu$-mesomolecules formation from the upper state  of the
hyperfine structure  $d\mu (F=3/2)$ starts earlier than the mesoatoms
thermolization. It explains an approximate constancy of the
$dd\mu$-mesomolecule formation rate in solid deuterium.

\vspace*{1.5cm}
\normalsize
\vspace*{0.5cm}
Highly effective resonant mechanism of the  $dd\mu$ mesomolecule formation
followed by the $dd\to ^3He+n$ or  $dd\to T+p$ nuclear synthesis becomes
possible in the gaseous and liquid deuterium due to the presence of the
oscillatory-rotary level $(\nu=1\;, \; K=1)$ in the  $dd\mu$ mesomolecule,
which has a small binding energy [1-3]. The matter is that the binding energy 
of the $dd\mu$ mesomolecule in this state, $(\epsilon\approx 2$~eV), is little
bit less than that which is needed to excite the oscillatory level $n=7$ of
the $[d\mu d,d]2e$ molecular complex, which arises when one of the 
nuclei of the $D_2$ molecule is substituted by the $dd\mu$ mesomolecule.
The lack of the energy is filled up by the thermal motion energy  of the  
$d\mu$-mesoatom thermolized in the matter, which collides with one
of the $D_2$-molecule deutrons, that allows to fulfill the condition of the  
resonant formation~[4]. 

The presence of two hyperfine structure levels with the total spin 
$F=3/2$ and $F=1/2$ in the  $d\mu$-mesoatoms (the energy difference
between them is equal to $\Delta E =0.0485$~eV) requires  a more detailed
investigation of the resonant mechanism. One should take into account
the fact that the  $dd\mu$-mesomolecule produced in the rotatory state $K=1$ 
with total nuclei spin equal to 1  via the resonant mechanism also has two
hyperfine structure levels  with the total spin 
$S=3/2$ and $S=1/2$, which have the energy difference $\Delta E= 0.242$~eV.
(the splitting of these levels caused by the mesomolecule rotation is
unimportant and can be neglected in the first approximation).

Analysing the time-depedence of the muon catalysis at different temperatures
one
can determine the rate of the mesomolecules formation from different 
states of the hyperfine structure of the $d\mu$-mesoatom as well as
the transition rate from upper level with the spin $F=3/2$ to lower level with
the spin $F=1/2$, which takes place in exchange collisions of the
$d\mu$-mesoatom with the deuterium nuclei~[5]. According to the
theoretical calculations [6] the rate of the resonant formation of the
$dd\mu$-molecule from the lowest state of the  $d\mu (F=1/2)$ mesoatom
hyperfine structure becomes smaller when the temperature of the gas matter or
liquid is decreased, since thermal energies of the thermolized 
$d_\mu$-mesoatoms are insufficient to fulfill the resonance condition.
At $T\approx 70^o$~K the resonance mechanism of the  
$dd\mu$-mesomolecules formation for the $d_\mu (F=1/2)$-state 
is almost completely ruled out and the 
$dd\mu$-mesomolecule formation from the $F=1/2$-state is solely induced by
the nonresonant mechanism, in which the mesomolecule binding energy is carried
off by atomic electron [7-11] (see Fig.~1). 

The probability of the nonresonant transition is about two order of magnitude
less than that for resonant one, and this mechanism leads to the formation of
the mesomolecules in the ground states with even rotatory numbers [2] and even
total nuclei spin $I=2,0$, and, in part, in the ground rotatory state
with $K=1,\;\; \nu=0$. 
The observed change of the ratio of the  $dd$-reaction channels,
$R=\frac{\Gamma(dd\to ^3He+n)}{\Gamma (dd\to T+p)}$~[12], which occurs in the
$\mu$-catalysis from the $F=1/2$ state when the temperature is decreased 
(that is due to the difference 
of this ratio for reactions starting from the rotatory state $K=1$ with total
nuclei spin $I=1$ and from the $S$-state with even rotatory numbers and spins
$I=0,\; I=2$), is considered as an additional experimental proof. 

As for the  process of the $dd\mu$-molecule formation from the  
$d_\mu (F=3/2)$-state, its rate weakly depends on the temperature and starts to
decrease at $T\simeq 26^o$K only. The temperature dependence of the molecule
formation rate measured in the dense gas and in liquid deuterium [13,14]
agreed well with the theoretical calculations [6] (see. Fig.~2). However, the
data on the $dd\mu$-mesomolecule formation rate in the solid hydrogen at
$T=3^o$~K [15] exceeds the theoretical expectations by more than order 
of magnitude. The detailed studies conducted on the same setup in
the solid deuterium and liquid  [16] in the temperature range of $5-50$~Š have
confirmed this result (see Fig. 2).   It turned out that
at  temperatures below $T\simeq 20^o$K the rate  of the resonant
formation of the $dd\mu$-mesomolecules from the $d\mu (F=3/2)$-state  does not
change practically. Analogously, in agreement with the theory,
the rate of the nonresonant formation from the  $d\mu (F=1/2)$-state does not
change too, and what is more important the rate of the transition between
hyperfine structure levels of the mesoatoms, $d_\mu (F=3/2)\to
d_\mu (F=1/2)$ ($\lambda_d\approx 3.1\cdot 10^7$c$^{-1}$) does not change too.
Thus, the paradoxical situation occurs.
It seems as if the  $d\mu$-mesoatoms in the hyperfine structure state
$F=3/2$ would conserve their energy and would not be thermolized with the
temperature decrease below $20^o$K. In this paper I would like to
point out that such phenomenon can be explained in a natural way in terms
of the type of energy loss by slow $d_\mu$-atoms in the crystal lattice and it
is almost independent on the lattice structure itself. 

Due to the electroneutrality an smallness of the $d_\mu$-mesoatoms 
their scattering in the crystal is analogous, in a large extent, to the
slow neutron scattering.
As it was shown by I.Ya.~Pomeranchuk in 1937 [17], the neutrons  at
sufficiently low temperatures occurs, in bulk, scatters elastically on 
whole crystal lattice and this process does not accompanied by the energy loss
(one could only regret that having obtained this result, I.Ya.~Pomeranchuk had
not predict the M\"ossbauer effect). As for inelastic collisions, they are
connected with the one-photon lattice excitation, and the mean free path,
$l_{in}$, is greater than elastic one $l_{el}$.  According to [17], for
identical nuclei (when the neutron wave length becomes smaller than the lattice
constant) one  gets
\begin{equation}
l_{in}=\frac{7}{8}l_{el} \left (\frac{k_B\Theta}{E}
\right )^3\; ,
\end{equation}
where $E$ is the neutron energy, $\Theta$ is the Debey temperature. At
$E <<k_B\Theta$ one has $l_{in} >>l_{el}$. Besides, if the length of the
neutron capture $l_c=\frac{1}{N\sigma_c}$ (where $\sigma_c$ is the capture
cross-section,  $N$ is the nuclei concentration) will be significantly smaller
than $l_{in}$, the neutron deceleration is ceased and they will be captured by
nuclei before their thermolization occurs. Analogous arguments can be applied
to the case of the $d\mu$-mesoatoms behaviour in crystal lattice. 

The cross-section of the elastic scattering of slow $d\mu$-atoms in the
hyperfine structure state  $F=3/2$ on deutons has the form [5]:
\begin{equation}
\sigma _{3/2\to 3/2}=4\pi
\left \{
\frac{1}{2} \lambda^2g
+\frac{1}{3} 
\left (
\frac{\lambda g+5\lambda u}{6}
\right )^2
+\frac{1}{6}
\left ( 
\frac{\lambda g+\lambda u}{3}
\right )^2
\right \},
\end{equation}
where $\lambda_g$ and $\lambda_u$ are the $d\mu$ scattering lengths in
potentials $V_g$ and $V_u$, which correspond to muon molecular orbits
$\Sigma_g$ and $\Sigma_u$ in the filed of two deutons
(where the first order corrections over the $m_\mu/M_d$ mass ratio due to the
motion nonadiabaticity are taken into account) [5]. The cross-section of the
$d\mu (F=3/2)\to d_\mu (F=1/2)$ transition due to collisions with muon exchange
is as follows
\begin{equation}
\sigma_{3/2\to 1/2} =\frac{\pi}{3}(\lambda_g-\lambda_u)^2
\frac{k_0}{k_1},
\end{equation}
where $k_1=\sqrt{\frac{M_dE}{\hbar}}$; 
$k_0=\sqrt{\frac{M_d(E+\Delta E)}{\hbar}}$, $E$ is the 
$d\mu (F=3/2)$ mesonucleus energy ,  $\Delta E=0.0485$~eV is the energy of the
hyperfine splitting in the  $d\mu$-atom.  
The rate of the transition into the lowest hyperfine structure state is [5]
\begin{equation}
\lambda_d=N\sigma_{3/2\to 1/2}\cdot v_1=\frac{\pi}{3}
(\lambda_g-\lambda_u)^2N\cdot v_0;
\end{equation}
$$
v_0=2\sqrt{\frac{\Delta E}{Md}}\simeq 3\cdot 10^5\mbox{cm}/\mbox{c}.
$$

Using the experimental value of $\lambda_d\simeq 3,1\cdot 10^7$ at
$N=4\cdot 25\cdot 10^{22}$cm$^{-3}$ one gets
$|\lambda_g-\lambda_u|=1,87\;a_\mu$
$\left (a_\mu=\frac{\hbar^2}{m_\mu e^2}=2,56\cdot 10^{-11}\mbox{cm}\right )$.
The estimates [5] show that scattering lengths $\lambda_g$ and $\lambda_u$ are
close to each other, and their average value is about $\bar \lambda=(5\div
6)a_\mu$. Thus, elastic cross-section  (2) can be rewritten in the following
form
\begin{equation}
\bar\sigma_{el}=4\pi \bar\lambda^2a^2_\mu.
\end{equation}
Mean free path, connected with $dd\mu$-mesomolecule formation, is equal to: 
\begin{equation}
l_{dd\mu}=
\frac{1}{N\sigma_{dd\mu}}=\frac{v}{\lambda_{dd\mu}},
\end{equation}
where $\sigma_{dd\mu}$ is the effective cross-section of mesomolecule
formation,  $v$ is the  $d\mu$-mesoatoms velocity, which under the 
thermolization condition is equal to 
$v=\left(
\frac{3k_BT}{M_d}\right )^{1/2}$. From the condition $l_{dd\mu}<l_{in}$ and 
equations (1) and  (5) one can estimate the temperature, at which the
mesomolecules formation starts early than mesoatoms thermolization:
\begin{equation}
T\leq 0.4  \Theta^{6/7}_d
\left (
\frac{M_d}{3k_B}\right )^{1/7}
\left (
\frac{\lambda_{dd\mu}}{\bar \lambda^2 a^2_\mu N}\right )^{2/7}
\end{equation}
For the experimental value  $\lambda dd\mu \simeq 2.27\cdot 10^6$1/s,
using  $\bar\lambda \sim 6a_\mu$ and assuming the Debey temperature for
deuterium equal to $\Theta_d=74^o$K, one gets
\begin{equation}
T\leq  12.5^o\mbox{K},
\end{equation}
that, even being roughly estimated, agrees well with the data [15,16]. 

Thus, in the solid deuterium the resonant formation of the
$dd\mu$-mesomolecules from upper level of the $d\mu$-mesoatom hyperfine
structure starts before the $d\mu$-atom thermolization. Namely this
phenomenon 
explains the independence of muon catalysis in sold deuterium on the
temperature. It should be noted, that ephythermal formation of the
$dd\mu$-mesomolecules has been observed earlier in the matter, composed of the
HD-molecules [18]. It was explained by the fact that the effective
cross-section of the $d\mu$-mesoatoms scattering on protons is small due to
the Ramsauer effect [19], and, thus, in the process of the $d\mu$-mesoatoms
deceleration their energies with particular probability have the values, which
correspond to the region of the resonant formation on higher oscillatory
states of the $[dd\mu,d]2e$ complex.

At same time,  the crystal lattice in solid deuterium slightly effects on the
rate of the resonant formation of the  $dt\mu$-mesomolecules due to the
difference in locations and widths of the resonant level for the  
$dt\mu$ and $dd\mu$ mesomolecules [20].

As I found out after this paper has been written, L.I.~Ponomarev, basing on the
experimental data analysis, has suggested the hypothesis that the
thermolization of the $d\mu$-atoms does not occur in solid deuterium    the
experimental data analysis. I sincere thank  L.I.~Ponomarev and G.G.~Semenchuk
for fruitful discussions.
This work was supported, in part, by the RFBR under grants
99-02-16558 and 00-15-96645.

\newpage
\section*{Figure captions}

\begin{description}
\item[1.] Molecular formation rates $\tilde\lambda_{1/2}$ and
$\tilde\lambda_{3/2}$ from all experiments performed at PSI and fits done in 
[21].
\item[2.]
Experimental dependence of the deuterium mesomolecule formation rate on
the temperature and spin state of the deuterium mesoatom. The data  points are
marked as follows: circles represent the experimental data [2],
square boxes -- the data from [15], triangle boxes -- the data from [14],
crossed points -- the data from  [13], solid line -- theoretical predictions
[6].
\end{description}

\end{document}